\documentclass[11pt]{amsart}

\begin{document}

\title[]{Black holes: the legacy of Hilbert's error}
\author{Leonard S. Abrams}
\address{\sl Lockheed Corporation, 4500 Park Granada Boulevard,
Calabasas, CA 91399, U.S.A., and 24345 Crestlawn Street, Woodland Hills, CA
91367, U.S.A. (present address).}

\thanks{Published in Can. J. Phys. {\bf 67} (1989) 919.}
\thanks{Received November 10, 1988}
\date{}

\begin{abstract}
The historical postulates for the point mass are shown to be
satisfied by an infinity of space-times, differing as to the
limiting acceleration of a radially approaching test particle.
Taking this limit to be infinite gives Schwarzschild's result,
which for a point mass at $x=y=z=0$ has $C(0+)=\alpha^2$, where
$\alpha=2m$ and $C(r)$ denotes the coefficient of the angular
terms in the polar form of the metric. Hilbert's derivation used
the variable $r^*=[C(r)]^{1/2}$, which transforms the coordinate
location of the point mass to $r^*_0=[C(0+)]^{1/2}$. For Hilbert,
however, $C$ was unknown, and thus could not be used to determine
$r^*_0$. Instead he asserted, in effect, that
$r^*=(x^2+y^2+z^2)^{1/2}$, which places the point mass at $r^*=0$.
Unfortunately, this differs from the value ($\alpha$) obtained by
substituting Schwarzschild's $C$ into the expression for $r^*_0$,
and since $C(0+)$ is a scalar invariant, it follows that Hilbert's
assertion is invalid. Owing to this error, in each spatial section
of Hilbert's space-time, the boundary ($r^*=\alpha$) corresponding
to $r=0$ is no longer a point, but a two-sphere. This renders his
space-time analytically extendible, and as shown by Kruskal and
Fronsdal, its maximal extension contains a black hole. Thus the
Kruskal-Fronsdal black hole is merely an artifact of Hilbert's
error.\par\vbox to 0.5 cm{}
Il existe un nombre infini d'espace-temps non \'equivalents pour
la masse punctuelle; ils diff\`erent les unes des autres quant \`a
l'acc\'el\'eration limite d' une particule d'essai s'approchant
radialement. En faisant cette limite infinie, on a l'espace-temps
inextensible de Schwarzschild, qui a, pour une masse punctuelle
\`a  $x=y=z=0$, $C(0+)=\alpha^2$, ou $\alpha=2m$ et $C(r)$
d\'esigne le coefficient des termes angulaires lorsque la
m\'etrique est \'ecrite en polaires sph\'eriques. Hilbert utilisait dans
sa d\'erivation la variable $r^*=[C(r)]^{1/2}$, qui transforme la
position de la masse punctuelle de $r^*_0=0$ \`a
$r^*_0=[C(0+)]^{1/2}$. Pour Hilbert cependant, $C$ \'etait une
inconnue, et il ne pouvait par cons\'equent l'utiliser pour
determiner $r^*_0$. Au lieu de cela, il affirmait en effet que
$r^*=(x^2+y^2+z^2)^{1/2}$, ce qui place la masse punctuelle \`a
$r^*=0$. Malhereusement, cette valeur diff\`ere de la
valeur ($\alpha$) obtenue en substituant le $C$ de Schwarzschild
dans l'expression de $r^*_0$; comme $C(0+)$ est une scalaire
invariant, il s'ensuit que l'affirmation de Hilbert est invalide.
Comme r\'esultat, dans chaque section spatiale de l'espace-temps
de Hilbert, la limite ($r^*=\alpha$) correspondant \`a $r=0$ n'est
plus un point mais une sph\`ere bidimensionnelle et par
cons\'equent pas une singularit\'e quasi r\'eguli\`ere. Cela rend
son espace-temps analytiquement extensible, et, comme l'ont
montr\'e Kruskal et Fronsdal, son extension maximale contient un
trou noir. Le trou noir Kruskal-Fronsdal n'est donc rien de plus
qu'un produit de l' erreur de Hilbert.
\end{abstract}
\maketitle

\section{Introduction}
Ever since Schwarzschild's 1916 derivation \cite{ref:Schw16a},
it has been accepted that the historical postulates used by him
to characterize the point-mass space-time do, in fact, lead
to a unique gravitational field. However, Schwarz\-schild's result
was obtained years before the concept of a `quasiregular'
singularity \cite{ref:ES77} had surfaced. As shown here, when the
general polar form of a static, spherically symmetric metric
\begin{equation}\label{1}
g_U(r)=A(r)dt^2-B(r)dr^2-C(r)d\Omega^2
\end{equation}
is substituted into Einstein's vacuum field equations and the
remaining historical postulates are imposed on the solution, there
results a one-parameter family of inequivalent, maximal
space-times, each of which has a quasiregular singularity at the
location of the point mass. Since each of these space-times
assigns a different number to the limiting value of a radially
approaching test particle's locally measured acceleration, it is
necessary to supplement the historical postulates by one that
fixes this limit. Taking it to be infinite gives Schwarzschild's
result
\begin{eqnarray}\nonumber
g_S(r)=(1-\alpha/R)dt^2-(1-\alpha/R)^{-1}dR^2-R^2d\Omega^2
\end{eqnarray}
where $\alpha=2m$ and $R=(r^3+\alpha^3)^{1/3}$.\par
Several months after Schwarzschild's paper was published, another
derivation of the point-mass space-time was given by Hilbert
\cite{ref:Hilbert17}. At the start he introduced the radial
coordinate $r^*=[C(r)]^{1/2}$, and then solved the vacuum field
equations for the two remaining unknowns, obtaining:
\begin{eqnarray}\nonumber
g_H(r^*)=(1-\alpha/r^*)dt^2-(1-\alpha/r^*)^{-1}dr^{*2}
-r^{*2}d\Omega^2
\end{eqnarray}
Of course, the use of $r^*=[C(r)]^{1/2}$ transforms the coordinate
location of the point mass from $r=0$ to $r^*=[C(0+)]^{1/2}$. As
can be seen from Schwarzschild's result, this is simply $\alpha$.
Hilbert, however, claimed that $r^*=(x^2+y^2+z^2)^{1/2}$, so that
the point mass represented by $g_H(r^*)$ was at $r^*=0$, and this
claim went unchallenged by his contemporaries. As a result of this
mistake, in each spatial section the locus $r^*=\alpha$ is not a
point but a two-sphere, and thus no longer constitutes a
quasiregular singularity of Hilbert's space-time $S_H$. As shown
by Kruskal \cite{ref:Kruskal60} and Fronsdal
\cite{ref:Fronsdal59}, with $r^*=\alpha$ having the character of a
two-sphere in the $t=$~constant hypersurfaces, $S_H$ is
analytically extendible to $r^*>0$, and the so-extended space-time
contains a black hole. It follows that the theoretical foundation
of spherical black holes is based on the 1916 error of
Hilbert.\par
It is the principal objective of this paper to show how Hilbert
came to make this mistake. It will also be shown that $S_H$ cannot
be used to represent the `exterior' of a collapsing star, and thus
that spherical black holes cannot be produced by gravitational
collapse.

\section{The space-time of a point mass}
Consider any static, time-symmetric and spherically symmetric
configuration ($U$) of matter and (or) energy. In his paper
\cite{ref:Schw16a} Schwarzschild regarded it as self evident that
the metric of such a configuration, when expressed in terms of
quasi-Cartesian coordinates ($t$, $x$, $y$, and $z$), was
necessarily of the form
\begin{eqnarray}\label{2}
g_U(x,y,z\mid0,0,0)&=F(r)dt^2-G(r)(dx^2+dy^2+dz^2)\\\nonumber
&-H(r)(xdx+ydy+zdz)^2
\end{eqnarray}
where
\begin{equation}\label{3}
r=(x^2+y^2+z^2)^{1/2}
\end{equation}
and
\begin{equation}\label{4}
F,G,H>0
\end{equation}
and where, without loss of generality, the center
\footnote[1]{Since a point mass is the only configuration under
consideration, it will be assumed here that only a single center
of symmetry is present.} \footnote[2]{The center is that
three-dimensional point having the property that rotations about
it leave the Newtonian description of $U$ unchanged. Note that the
very concept of spherical symmetry presupposes that the center of
symmetry is a three-dimensional point in each spatial section
\cite{ref:Demianski85}.} of the spherical symmetry has been taken
at $x=y=z=0$ for all $t$; this is signified by the notation
`$\mid0,0,0$' in the argument of $g_U$. (The generality of
(\ref{2}) for such configurations was subsequently established by
Eiesland \cite{ref:Eiesland25}.)\par
In polar coordinates ($r$, $\theta$, and $\phi$), where
$x=r\sin\theta\cos\phi$, etc., the above expression of $g_U$
becomes
\begin{equation}\label{5}
g_U(r\mid 0)=A(r)dt^2-B(r)dr^2-C(r)d\Omega^2
\end{equation}
where
\begin{equation}\label{6}
d\Omega^2=d\theta^2+d\phi^2\sin^2\theta
\end{equation}
and from (\ref{4}) and the tensor transformation law, it follows
that
\begin{equation}\label{7}
A,B,C>0
\end{equation}
(Analogously, the `$\mid0$' in the argument of $g_U$ in (\ref{5})
signifies that the center of symmetry of (\ref{5}) is at $r=0$ for
all $t$.)\par
For later purposes, we note that apart from spatial rotations
about the origin, which are of no relevance to what follows, the
only transformations $T$ that leave the structure of (\ref{5})
unchanged are of the form
\begin{eqnarray}\label{8}
t=k\bar{t}+b,&k\neq0,~b~constants\\\label{9}
r=h(\bar{r}),&h\in C^1
\end{eqnarray}
Since $C(r)$ in (\ref{5}) is readily seen to be a scalar under
$T$, it follows that:\par
\vbox to 0.3 cm{}
\noindent
THEOREM:{\it The value of $C$ at a specified event $P$} ({\it
this will be denoted by} `$C[P]$') {\it is an invariant of the
space-time associated with a given $U$.}\par
\vbox to 0.3 cm{}
\noindent
Consequently, two space-times $S_1$ and $S_2$ having metrics of
the form (\ref{5}) for which $C_1[P]\neq C_2[P]$ are necessarily
inequivalent. {\it A fortiori},\par
\vbox to 0.3 cm{}
\noindent
COROLLARY:{\it If two space-times $S_1$ and $S_2$ have metrics of
the form} (\ref{5}){\it , and if $\lim C_1[P]\neq\lim C_2[P]$ as $P$
approaches the center of symmetry, then $S_1$ and $S_2$ are
inequivalent.}\par
\vbox to 0.3 cm{}
Consider now the particular $U$ consisting of a single uncharged,
nonrotating, nonradiating point mass (whose Newtonian
gravitational mass will henceforth be denoted by `$m$').
Historically, the conditions regarded as distinguishing the
space-time ($M_U$,~$g_U$) of this $U$ from those of all others
where originally formulated by Einstein \cite{ref:Einstein15}, and
together with those implicit in that formulation were enumerated
by Finkelstein \cite{ref:Finkelstein58}. The expression on the
right-hand side of (\ref{5}), together with (\ref{6}) and
(\ref{7}), already satisfies the static- and spherical-symmetry
requirements of Finkelstein's list, as well as those requiring a
Lorentz signature and a global-time coordinate. As shown in
Appendix A, the further requirements that (\ref{5}) be analytic,
satisfy the vacuum field equations, and be asymptotically flat
reduce (\ref{5}) to
\begin{equation}\label{10}
g_{PM}(r\mid0)=(1-\alpha/C^{1/2})dt^2
-\frac{{C'}^2}{4C(1-\alpha/C^{1/2})}dr^2-Cd\Omega^2
\end{equation}
where $C(r)$ is any analytic function of the $r$ in (\ref{3})
having the following three properties:
\begin{eqnarray}\label{11}
C(0+)\equiv\beta^2\geq\alpha^2\\\label{12}
C'(r)>0,~r>0\\\label{13}
C(r)/r^2\rightarrow1~as~r\rightarrow\infty
\end{eqnarray}
(the prime denotes differentiation with respect to $r$). As shown
there, all such metrics comply with the final requirement of being
analytically inextendible to $r=0$.\par
Since the center of symmetry of (\ref{5}) is located at $r=0$, the
same is true of the center of symmetry of (\ref{10}), whence it
follows from the corollary that two space-times defined on
$M_0:r>0$ and having $g_{PM}$ as their metric are inequivalent if
$C_1(0+)\neq C_2(0+)$. This, together with (\ref{11}), shows that
there is a one-parameter family of inequivalent space-times
(distinguished by their value of $C(0+)$), which satisfy all the
historical postulates for the point mass. That is:
\newpage
\begin{quote}
The historical postulates for the point mass do not result in a
unique space-time. To obtain such a space-time, it is necessary to
supplement those postulates by one which fixes the value of
$C(0+)$.\footnote[3]{The fact that a boundary condition for
$r\rightarrow 0$ is just as necessary as the one for
$r\rightarrow\infty$ was first realized by Brillouin
\cite{ref:Brillouin70}. Although Abrams' work (see ref.
\cite{ref:Abrams79}) contained a derivation of the general
point-mass metric similar to that given here in Appendix A, it made use
of an invalid argument (see ref. \cite{ref:Abrams80}) to prove
that $C(0+)=\alpha^2$, and thus did not turn up the fact that an
additional postulate was required.}
\end{quote}
\noindent To determine this supplementary postulate, it is
necessary to relate $C(0+)$ to some physical property of the point
mass. To this end, consider the motion of an uncharged,
nonspinning, nonradiating test particle as it approaches the point
mass along a radial geodesic. As shown by Doughty
\cite{ref:Doughty81a}, the particle ``locally measured''
acceleration (i.e., its acceleration measured by a nearby observer
whose position is fixed with respect to the point mass (see
Appendix B) is given by:
\begin{equation}\label{14}
a=\frac{(-g_{rr})^{1/2}(-g^{rr})|g_{00,r}|}{2g_{00}}
=\frac{\alpha}{2C(1-\alpha/C^{1/2})^{1/2}}
\end{equation}
which approaches
\begin{equation}\label{15}
a_0=\frac{\alpha}{2\beta^{3/2}(\beta-\alpha)^{1/2}}
\end{equation}
as $r\rightarrow 0$. Consequently, for a given value of $m$, the
value of $C(0+)$ is determined by the limiting value of a
test-particle's locally measured acceleration as it approaches the
point mass along a radial path.\par
Since there are no experimental data concerning $a_0$, it is
necessary to choose its value on the basis of theoretical
arguments. One such argument is that in the corresponding
Newtonian situation $a_0$ is infinite. Although one would not
expect strict quantitative agreement (in the sense of equal values
of $a$ at equal values of proper distance), the fact that the
curvature invariant $f=R_{ijkm}R^{ijkm}$ remains finite as
$r\rightarrow 0$ indicates that the relativistic case differs from
the Newtonian one only in degree, not in kind. Accordingly, since
there are no ``degrees'' associated with an infinite limit, the
supplementary postulate is taken to be:
\begin{quote}
({\it i}) the limiting value of a neutral nonspinning nonradiating
test particle's locally measured acceleration as it approaches the
point mass along a radial geodesic is infinite.
\end{quote}
Examination of (\ref{15}) shows that the only values of $\beta$
that make $a_0$ infinite are $0$ and $\alpha$, and of these only
the latter satisfies (\ref{11}). Hence, the result of postulate
({\it i}) is
\begin{equation}\label{16}
C(0+)=\alpha^2
\end{equation}
The simplest choice of analytic $C$ satisfying (\ref{12}),
(\ref{13}), and (\ref{16}) is
\begin{equation}\label{17}
C_B(r)=(r+\alpha)^2
\end{equation}
which when substituted into (\ref{10}) results in the following
form, first used by Brillouin \cite{ref:Brillouin22}, for the
metric of a point mass at $r=0$:
\begin{equation}\label{18}
g_B(r\mid 0)=\frac{r}{r+\alpha}dt^2-\frac{r+\alpha}{r}dr^2
-(r+\alpha)^2d\Omega^2
\end{equation}
The metric obtained by Schwarzschild corresponds to the choice
\begin{equation}\label{19}
C_S(r)=(r^3+\alpha^3)^{2/3}
\end{equation}
which satisfies (\ref{12}), (\ref{13}), and (\ref{16}) by
inspection.\par
It should be noted that all space-times $[M_0,~g_{PM}(r\mid 0)]$
arising from (\ref{10}) via an analytic $C$ satisfying
(\ref{12}), (\ref{13}), and (\ref{16}) are equivalent, being
related via $C^{\omega}$ global diffeomorphisms of the form
\begin{equation}\label{20}
C_i(r)=C_j(\bar{r})
\end{equation}
and being globally homeomorphic and maximal, as well as having the
same singularity at $r=\bar{r}=0$. Consequently, the
space-time $S_S=[M_0,~g_B(r\mid~0)]$ will henceforth be termed
``Schwarzschild's'', since it is equivalent to the
$[M_0,~g_S(r\mid 0)]$ actually obtained by him, and its use
simplifies the subsequent discussion.

\section{Transformation to Flamm's form}
Let us relabel the events of $M_0$ by adding $\alpha$ to their $r$
values, and denote the new ``radial'' coordinate by $\bar{r}$;
then
\begin{eqnarray}\label{21}
\bar{r}=r+\alpha,~r>0
\end{eqnarray}
This transforms $M_0$ to $\bar{M}_{\alpha}$, where
\begin{equation}\label{22}
\bar{M}_{\alpha}:\bar{r}>\alpha
\end{equation}
tranforms the location of the point mass at $r=0$ to
$\bar{r}=\alpha$, and transforms $g_B(r\mid 0)$ to
\begin{equation}\label{23}
g_F({\bar{r}}\mid\alpha)=\frac{\bar{r}-\alpha}{\bar{r}}dt^2
-\frac{\bar{r}}{\bar{r}-\alpha}d\bar{r}^2
-\bar{r}^2d\Omega^2
\end{equation}
termed the Flamm metric since Flamm \cite{ref:Flamm16} was the
first to use this form to represent a point mass at
$r=\alpha$.\par
(To avoid misunderstanding, note that the statement: ``The point
mass is at $r=0$,'' in connection with (\ref{18}), and ``The point
mass is at $\bar{r}=\alpha$'', in connection with (\ref{23}), do not
mean that these $r$ or $\bar{r}$ values are well-defined
quantities, let alone that events having such values are part of
the associated space-times. What they do mean is that, in the case
of (\ref{18}), the proper distance from the point mass to an event
in $M_0$ with coordinate $r$ tends to $0$ as $r\rightarrow 0$; and
in the case of (\ref{23}), that the proper distance from the point
mass to an event in $\bar{M}_{\alpha}$ {\it with coordinate}
$\bar{r}$ tends to $0$ as $\bar{r}\downarrow\alpha$.)\par
Since physics is not changed by such a relabelling, it follows
that the Flamm space-time:
\begin{equation}\label{24}
S_F=[\bar{M}_{\alpha},~g_F(\bar{r}\mid\alpha)]
\end{equation}
represents the identical physical situation as $S_S$, namely, a
single uncharged, nonrotating, nonradiating point mass. Thus, both
$S_S$ and $S_F$ have the same singularity structure, i.e., a
quasiregular singularity at the location of the point mass, and thus
$S_F$ is likewise analytically inextendible to that location, now
denoted by $\bar{r}=\alpha$.\par
In passing, note that the fact that the coefficient of $d\Omega^2$
in (\ref{23}) tends to $0$ as $\bar{r}\downarrow 0$ does not
contradict (\ref{11}), since the $\bar{r}$ in (\ref{23}) is not
the $r$ in (\ref{3}).

\section{Hilbert's derivation of the point-mass metric}
About 10 months after the publication of ref.
\cite{ref:Schw16a}, Hilbert \cite{ref:Hilbert17} presented another
derivation of the point-mass metric. Although Hilbert's starting
point was the same expression (\ref{5}) used by Schwarzschild, he
immediately reduced the number of unknowns to two by introducing a
new ``radial'' coordinate $r^*$, defined by
\begin{equation}\label{25}
r^*=[C(r)]^{1/2}
\end{equation}
This transforms (\ref{5}) to
\begin{equation}\label{26}
g^*_U(r^*\mid 0^*)=A^*(r^*)dt^2-B^*(r^*)dr^{*2}-r^{*2}d\Omega^2
\end{equation}
where from (\ref{7})
\begin{equation}\label{27}
A^*,B^*>0
\end{equation}
and $0^*$ denotes the value of $r^*$ at the location of the point
mass, which from (\ref{25}) is given by
\begin{equation}\label{28}
0^*=[C(0+)]^{1/2}
\end{equation}
Solving the resulting vacuum field equations by means of a
variational principle, Hilbert arrived at the following expression
for the point-mass metric:
\begin{equation}\label{29}
g_F(r^*\mid
0^*)=\frac{r^*-\alpha}{r^*}dt^2-\frac{r^*}{r^*-\alpha}dr^{*2}
-r^{*2}d\Omega^2
\end{equation}
which by inspection is well defined on
$M^*_{\alpha}:r^*>\alpha$.\par
However, there are two problems connected with the use of
(\ref{25}), which Hilbert evidently overlooked. The first of these
is that (\ref{25}) involves a loss of generality \cite{ref:KB80};
for example, if $U$ is such that the $C$ in (\ref{5}) is constant
\cite{ref:Nariai50}, then the use of (\ref{25}) will make it
impossible to determine the metric for $U$. Unfortunately (from
the stand-point of subsequent developments), this flaw had no
impact on Hilbert's derivation, since as seen in Sect. 2, for the
point-mass metric all $C(r)$ permissible in (\ref{5}) are strictly
monotonic, so that (\ref{25}) is in fact a diffeomorphism for such
a $U$.\par
The second is that the use of (\ref{25}) destroys information;
{\it once it is applied, it becomes impossible to determine the
relationship between $r^*$ and $r$ (and thus to find the value of $0^*$),
since at this point in the derivation the function $C(r)$
in (\ref{25}) is unknown, and there is no way to determine what it
is (or even its value as $r\rightarrow 0$) from the resulting
$A^*$ and $B^*$.}\par
This being the case, how {\it did} Hilbert arrive at the value of
$0^*$? The answer is as follows: by assuming that for all $U$, the
triplet ($r^*$, $\theta$, and $\phi$) can be regarded as polar
coordinates just as validly as ($r$, $\theta$, and $\phi$), so
that $r^*=(x^2+y^2+z^2)^{1/2}$. From this assumption, together
with the fact that the location of the point mass described by
(\ref{5}) is given by $x=y=z=0$, it follows at once that even without
knowing $C(r)$, the $r^*$ position of the point mass described by
(\ref{26}) is given by $r^*=0$, and so Hilbert proclaimed.\par
While it is true that there are a number of $U$ for which
Hilbert's assumption is valid, the point mass is not one of them,
as shown in Sect. 2. For such a $U$ all $C$ in (\ref{5}), and thus
in (\ref{25}), have the property that $C(0+)=\alpha^2$, whence
(\ref{28}) shows that $0^*=\alpha>0$. Thus, as the notation on the
left-hand side of (\ref{29}) indicates, the right-hand side is
simply Flamm's metric for a point mass at $r^*=\alpha$.\par
Henceforth, to avoid circumlocution the right-hand side of
(\ref{29}) together with the assumption that the $r^*$ appearing
therein is related to the $x$, $y$, and $z$ in (\ref{2}) via
(\ref{3}), and thus that the point mass described by (\ref{29}) is
at $r^*=0$, will be termed ``Hilbert's'' metric \footnote[4]{In
ref. \cite{ref:Abrams79}, this metric was designated
``Droste-Weyl''. The reason was that both Droste
\cite{ref:Droste17} in 1916 and Weyl \cite{ref:Weyl17} in 1917 had
derived metrics having the {\it form} of $g_H$, whereas my copy of
Hilbert's paper was dated 1924, so that it appeared that Droste
and Weyl had been the first to obtain $g_H$. However, I
subsequently discovered that Hilbert's paper had originally been
published in 1916, and that neither Droste nor Weyl had claimed
that the $r$ coordinate appearing in their metrics was
$(x^2+y^2+z^2)^{1/2}$. Thus, Hilbert alone was responsible for the
error.}, and denoted by $g_H$:
\begin{equation}\label{30}
g_H(r^*\mid 0)
=\frac{r^*-\alpha}{r^*}dt^2-\frac{r^*}{r^*-\alpha}dr^{*2}
-r^{*2}d\Omega^2
\end{equation}
while Hilbert's space-time will be denoted by $S_H$
\begin{equation}\label{31}
S_H=[M^*_{\alpha},~g_H(r^*\mid 0)]
\end{equation}

\section{Extendibility of Hilbert's space-time}
Consider the circle $\gamma(\epsilon)$ mentioned at the end of
Appendix A. In $S_S$, its description is given by $t=t_0$,
$\theta=\pi/2$, and $r=\epsilon$, where $t_0$ is an arbitrary
constant, independent of $\epsilon$. Since $S_S$ is diffeomorphic
to $S_H$ via $T_{\alpha}:M_0\rightarrow M^*_{\alpha}$ by
$r^*=r+\alpha$, the description of $\gamma(\epsilon)$ in $S_H$ can
be obtained by applying $T_\alpha$ to its description in $S_S$.
This gives  $t=t_0$, $\theta=\pi/2$, and $r^*=\epsilon+\alpha$. As
seen from (\ref{30}), its proper circumference is still
$2\pi(\epsilon+\alpha)$, which as in Appendix A tends to
$2\pi\alpha$ as $\epsilon\downarrow 0$. Now, however, in contrast
to the situation in Appendix A, this no longer gives rise to a
violation of elementary flatness at $T_\alpha(0)=\alpha$ in $S_H$
because, thanks to the assumption that gave birth to (\ref{30}),
$\gamma(\epsilon)$ no longer shrinks down to a point as
$\epsilon\downarrow 0$, but instead simply approaches the circle
$r^*=\alpha$. Thus, {\it Hilbert's erroneous (for a point mass)
assumption results in the disappearance of the quasiregular
singularity that is present in Flamm's space-time at
$\bar{r}=\alpha$.} Since it is well known that there are no
curvature-type singularities of $g_H$ at $r^*=\alpha$, it follows
that there are no singularities of any kind there, so that $S_H$
is analytically extendible to $r^*=\alpha$, and as Kruskal and
Fronsdal have shown, all the way to $r^*>0$.

\section{Inequivalence of Schwarzschild and Hilbert space-times}
If two space-times are to be equivalent, it is certainly necessary
that they be isometric i.e., that there exists a diffeomorphism
from one to the other that carries the metric of one into the
metric of the other. And since the presence of singularities of
the manifold geometry is unaffected by diffeomorphisms, it is also
necessary that equivalent space-times have the same ``singularity
structure'', i.e., the same singularities as one approaches
corresponding boundary points. Now, $S_S$ and $S_H$ are isometric
under $T_\alpha$, but as shown in the preceding section, $S_H$ has
no singularity corresponding to the quasiregular singularity at
$r=0$ in $S_S$. Consequently, $S_S$ and $S_H$ are inequivalent.
Since it was shown in Sect. 2 that $S_S$ is the space-time of a
point mass, it follows that $S_H$ and its analytic extension
($S_{K-F}$) are not.
\section{The Kruskal-Fronsdal black hole is unnecessary}
Consider any phenomenon supposedly involving a single
Kruskal-Fronsdal (K-F) black hole (e.g., x-ray spectra from
accreting gas). Because of the infinite red shift at the surface
($r^*=\alpha$) of the hole, all that we can ever know of this
phenomenon must arise from information originating outside the
hole. But the space-time ``exterior'' to the hole is $S_H$, which
in turn is diffeomorphic to $S_S$. That is to say, everything that
takes place outside the hole would occur in the {\it identical}
fashion if the entire space-time ($S_{K-F}$) were replaced by
$S_S$; it is impossible to determine which space-time is
``really'' present. Thus, any observations that are explicable by
postulating the presence of a K-F black hole are equally well
explained by postulating the presence of a Schwarzschild point
mass at the geometrical center of the black-hole's surface.
Consequently, there is no need to involve a K-F black hole to
explain {\it any} set of observations, Schwarzschild's ``black
point'' will do an equally effective job.\par
Whether this observational equivalence extends to the case of two
or more black holes (vis-a-vis two or more point masses) is
unclear, but in view of the remarks of the next two sections it
does not seem worthwhile to pursue the matter.

\section{The K-F black hole is unproducible}
The valid proofs of Birkhoff's theorem (e.g. refs.
\cite{ref:Eiesland25} and \cite{ref:Bonnor62}) show only that any
spherically symmetric solution of the vacuum-field equations can
be given the form of $g_H$, but say nothing as to the relationship of
the radial coordinate of the transformed metric and the underlying
quasi-Cartesian $x$, $y$, and $z$. As emphasized by Brans
\cite{ref:Brans65}, until this relationship is known the metric is
undefined. Thus, Birkhoff's theorem cannot be used to justify the
claim that the metric exterior to a spherically symmetric star is
identical to that of a point mass.\par
However, the characteristics of the space-time exterior to a
spherically symmetric, uncharged, nonrotating, nonradiating star
are nearly the same as those of a point mass, the only difference
being that for the star the postulates relating to the behavior of
the space-time at $r=0$ are no longer applicable.
Consequently, to determine the exterior metric for such a star all
that is necessary is to impose the remaining postulates for the
point mass on (\ref{5}). As can be seen by inspection of Appendix
A, doing so changes nothing up through (\ref{A16}), while (\ref{A17})
must be replaced by
\begin{equation}\label{32}
C(r_b)\geq\alpha^2
\end{equation}
where $r_b$ denotes the $r$ coordinate of the star's boundary.
(The precise value of $C(r_b)$ consistent with (\ref{32}) is
determined by the junction conditions.) Consequently, the exterior
metric of such a star is given by $g_{PM}(r\mid 0)$ in (\ref{10}),
but with $C$ now satisfying (\ref{12}) for $r>r_b$, (\ref{13}),
and (\ref{32}). That is to say, the exterior metric has the same
functional form as the metric of a point mass located at the
star's center, but the values of the parameters appearing therein
are different. An example of such a situation can be found in
ref.~\cite{ref:Schw16b}.\par
In the case of collapse to a point, this distinction
ultimately vanishes, so that the appropriate space-time for the
exterior of a star undergoing catastrophic collapse tends to the
($S_S$) of a point mass. While the precise details of the approach
to $S_S$ will vary from case to case, it is already clear
from the form of $g_{PM}$, together with (\ref{13}) and
(\ref{32}), that $A(r)>0$ for all $r>r_b$, and thus, that no black
hole ever forms in the exterior of the star, no matter how far the
collapse proceeds.\par
Thus, the correction of Hilbert's error not only eliminates the
point mass as a possible source of K-F black holes, but
simultaneously deprives them of the only mechanism for their
production.

\section{The K-F black hole is unreal}
Although it was shown in Sect. 6 that $S_{K-F}$ does not represent
the space-time of a point mass, it might still be hoped that it
represents some other configuration of matter and (or) energy, and
is thus of physical significance in its own right.\par
However, this is not the case. Since the energy-momentum tensor
vanishes everywhere in $S_{K-F}$, the only possible locations of
its sources are at its singularities. These are at $r^*=0$, or in
terms of Kruskal's $u$ and $v$, at $v^2-u^2=1$. As is easily seen,
these loci are spacelike, whereas those of real matter or
radiation are timelike or null, respectively. Consequently, it is
impossible for $S_{K-F}$ to represent any real configuration of
matter and (or) energy, i.e., $S_{K-F}$ is physically unreal, and
thus so is the K-F black hole.\par
While this disposes of the reality of $S_{K-F}$, $S_{K-F}$
possesses another property worth mentioning in connection with
other types of black holes, namely, the presence at $r^*=\alpha$
of an interior ``surface of infinite acceleration'' (at which the
locally measured acceleration of test particles becomes infinite:
see ref. \cite{ref:Rindler60}), despite the fact that there are
neither matter singularities nor geometrical singularities at that
surface. This existence of such a surface interior to the
space-time is a direct result of Hilbert's assumption regarding
$r^*$, since it was this that transferred the ``Cheshire cat''
(the point mass) to a distant point, while leaving the ``grin''
(the infinite acceleration of test particles) intact at
$r^*=\alpha$.

\section{Conclusions}
We summarize the result of the preceding sections as follows. The
K-F black hole is the result of a mathematically invalid
assumption, explains nothing that is not equally well explained by
$S_S$, cannot be generated by any known process, and is physically
unreal. Clearly, it is time to relegate it to the same museum that
holds the phlogiston theory of heat, the flat earth, and other
will-o'-the-wisps of physics.\par
Consider next the family ($\mathcal{S}$) of black holes obtained by
analytic extension of metrics, which, for certain values
($\beta_0$), of their parameters ($\beta$), reduce to $g_H$ (e.g.,
Reissner-Nordstr\"om, Kerr, Kerr-Newman, etc.). Some, like
Reissner-Nordstr\"om's solution for the point charge, were likewise
derived from a set of postulates characterizing the specified
matter and (or) energy configuration, and thus their derivations
must be analyzed to determine whether an error such as that made
by Hilbert was committed.\footnote[5]{A paper proving that this is
indeed the case for the point-charge metric is shortly to be
submitted by the author.} Others were simply ``discovered'', and
their sources sought afterwards, so it is impossible to determine
whether they are based on an invalid assumption. However, the fact
that all such metrics are regarded as reducing to $g_H$ when
$\beta=\beta_0$ (see ref. \cite{ref:Hernandez68}, \cite{ref:TS72},
and \cite{ref:LPPT75}) shows that it is tacitly assumed that the
radial coordinate appearing in these metrics is equal to
$(x^2+y^2+z^2)^{1/2}$, and Sect.~4 shows that this assumption is
invalid when $\beta=\beta_0$. Moreover, each of the associated
black-hole spacetimes bears the telltale stigma (an interior
surface of infinite acceleration, see ref. \cite{ref:Doughty81b})
associated with the transfer of boundary behavior to interior
events that was shown in the previous section to be a consequence
of precisely that assumption for the case of K-F black holes.
Accordingly, members of $\mathcal{S}$ are highly suspect.\par
Finally, there remain those black holes that have likewise simply
been discovered, but whose exterior metrics do not reduce to $g_H$
for any values of their parameters (e.g., the ``toroidal'' black
holes described in ref \cite{ref:GH82}). Their status awaits an
investigation of the reasonableness of their sources.

\section*{Acknowledgments}
It is a pleasure to acknowledge a number of helpful conversations
with B. O'Neill and R. Greene, and with M. Morris, as well as the
critiques of the author's earlier paper on this subject by R.
Gautreau and C. Will.

\section*{Erratum \cite{ref:Abrams96}}
Section 6 should be replaced by the following paragraph:
\section*{6. Inequivalence of Schwarzschild and Hilbert universes}
By inspection, $S_S$ and $S_H$ are isometric via $T_\alpha$ and
thus equivalent. However, it was shown above that due to the
difference in the topology of their boundaries, they are
associated with different singularity structures. Thus, the
universes ($U_S$ and $U_H$) corresponding to $S_S$ and $S_H$ (with
their indicated boundaries) are inequivalent (cf. Abrams, L.S.,
Physica A, {\bf 227} (1996) 131). Since it follows from Sect. 2
that $U_S$ is the universe of a point-mass, then a fortiori $U_H$
is not. For the same reason, this last is also true of $U_{KF}$,
the universe corresponding to the maximal analytic extension
($S_{KF}$) of $S_H$ found by Kruskal and Fronsdal.\par\bigskip

Section 7, line 12 should read: black hole's ``surface''.\par
\vbox to 1.0 cm{}
\bibliographystyle{amsplain}

\numberwithin{equation}{section}
\appendix
\section{Derivation of point-mass metric}

Substituting (\ref{5}) into Dingle's expressions (see ref.
\cite{ref:Tolman34a}) for $T^i_j$ gives
\begin{eqnarray}\label{A1}
-8\pi T^1_1\equiv -\frac{1}{C}+\frac{{C'}^2}{4BC^2}
+\frac{A'B'}{2ABC}=0\\\nonumber
-8\pi T^2_2\equiv\frac{C''}{2BC}+\frac{A''}{2AB}
-\frac{{C'}^2}{4BC^2}-\frac{B'C'}{4B^2C}-\frac{{A'}^2}{4A^2B}\\\label{A2}
-\frac{A'B'}{4AB^2}+\frac{A'C'}{4ABC}=0\\\label{A3}
T^3_3=T^2_2=0\\\label{A4}
-8\pi T^4_4\equiv\frac{C''}{BC}-\frac{1}{C}-\frac{B'C'}{2B^2C}
-\frac{{C'}^2}{4BC^2}=0
\end{eqnarray}
with all other $T^i_j$ identically zero. (Here and afterwards a
superscript prime denotes differentiation with respect to
$r$.)\par
Subtracting (\ref{A1}) from (\ref{A4}) and multiplying the result
by $BC$ (nonzero because of (\ref{7})) gives
\begin{equation}\label{A5}
C''-\frac{C'}{2}[\ln(ABC)]'=0
\end{equation}
Since $C'=0$ would (in view of (\ref{7})) reduce (\ref{A1}) to
$-1=0$, it follows that $C'\neq 0$, hence dividing (\ref{A5}) by
$C'$ gives
\begin{equation}\label{A6}
\frac{-2C''}{C'}-[\ln(ABC)]'=0
\end{equation}
which integrates at once to
\begin{equation}\label{A7}
{C'}^2=JABC
\end{equation}
with $J$ a constant. Since $A$, $B$, and $C>0$ and $C'\neq 0$, it
follows that $J>0$, and thus $C'$ never vanishes for $r>0$.\par
Solving (\ref{A7}) for $B$ and substituting the result into
(\ref{A1}) gives, after cancelling some nonzero factors,
\begin{equation}\label{A8}
\frac{-1}{C}+\frac{JA}{4C}+\frac{JA'}{2C'}=0
\end{equation}
whence either
\begin{equation}\label{A9}
\frac{C'}{C}=\frac{2JA'}{4-JA}~or~A=\frac{4}{J}
\end{equation}
The second alternative results in no gravitational force whatever
on a distant test particle (as seen from ref. \cite{ref:Tolman34b},
the gravitational acceleration of such a particle in the field of
(\ref{5}) is $-\psi'/2$, where $\psi=A-1$; thus, $A=$~constant
would result in zero acceleration), and is therefore ruled out.
The first integrates to
\begin{equation}\label{A10}
C(JA/4-1)^2=K_0\equiv\alpha^2>0
\end{equation}
the positivity of the constant $K_0$ being a consequence of that
of $C$ and of the non-negativity of $(JA/4-1)^2$. Solving
(\ref{A10}) for $A$ (and without loss of generality, choosing
$\alpha>0$) gives
\begin{equation}\label{A11}
A=\frac{4}{J}\left(1\pm\frac{\alpha}{C^{1/2}}\right)
\end{equation}
which upon substitution into (\ref{A7}) yields
\begin{equation}\label{A12}
B=\frac{{C'}^2}{4C(1\pm\alpha/C^{1/2})}
\end{equation}
Substituting $A$ and $B$ from (\ref{A11}) and (\ref{A12}) into
(\ref{A2}) shows that the latter is satisfied identically for
arbitrary $C$.\par
Now, the requirement that (\ref{5}) be spatially asymptotically
flat necessitates that
\begin{equation}\label{A13}
\frac{C}{r^2}\rightarrow 1~as~r\rightarrow\infty
\end{equation}

\noindent From this and from the fact that $C'$ cannot be zero it follows
that
\begin{equation}\label{A14}
C'>0~for~r>0
\end{equation}
Moreover, from (\ref{A13}) and (\ref{A11}) we see that
\begin{equation}\label{A15}
A\rightarrow 4/J~as~r\rightarrow\infty
\end{equation}
whence asymptotic flatness requires that $J=4$. This reduces
(\ref{A11}) to
\begin{equation}\label{A16}
A=1-\frac{\alpha}{C^{1/2}}
\end{equation}
the choice of the minus sign being compelled by the fact that the
gravitational force on a distant test particle must be attractive
(as noted above, the gravitational acceleration of a distant test
particle is $-\psi'/2$, where $\psi=A-1=\pm\alpha/C^{1/2}$; this
acceleration will only be attractive if the lower sign is chosen,
since both $C$ and $C'$ are positive). From (\ref{7}), (\ref{A14}),
and (\ref{A16}) it follows at once that
\begin{equation}\label{A17}
C(0+)\geq\alpha^2>0
\end{equation}\par
Hence we conclude that the most general triplet satisfying the
historical postulates for the point mass is of the form
\begin{eqnarray}\label{A18}
A=1-\alpha/C^{1/2},~\alpha>0\\\label{A19}
B=\frac{{C'}^2}{4C(1-\alpha/C^{1/2})}
\end{eqnarray}
where $C$ is an analytic function of $r=(x^2+y^2+z^2)^{1/2}$
having the following properties:
\begin{eqnarray}\label{A20}
C(0+)\equiv\beta^2\geq\alpha^2>0\\\label{A21}
C'(r)>0,~for~r>0\\\label{A22}
C(r)/r^2\rightarrow1~as~r\rightarrow\infty
\end{eqnarray}\par
Finally, consider the circle $\gamma(\epsilon):~t=t_0$
(a constant), $\theta=\pi/2$, and $r=\epsilon$. Inspection of
(\ref{10}) shows that the proper circumference of
$\gamma(\epsilon)$ is $2\pi[C(\epsilon)]^{1/2}$, which by
(\ref{A20}) tends to $2\pi\beta\geq2\pi\alpha>0$ as
$\epsilon\downarrow0$. Moreover, inspection of (\ref{10}) and
(\ref{A19}) shows that the proper radius \footnote[6]{If the
boundary $r=0$ in the spatial section is a point, which is the
case by virtue of the point mass being located there.} of
$\gamma(\epsilon)$ is
\begin{eqnarray}\label{A23}
R_P(\epsilon)=\int^\epsilon_{0-}[B(r)]^{1/2}dr\\\nonumber
=\int^\epsilon_{0-}
\frac{C'dr}{2C^{1/2}(1-\alpha/C^{1/2})^{1/2}}\\\label{A24}
=\int^{[C(\epsilon)]^{1/2}}_\beta
\left[\frac{u}{u-\alpha}\right]^{1/2}du
\end{eqnarray}
Since $C(0+)$ is finite, and $C(r)$ monotonic and analytic for
$r>0$, these last two properties are also true of
$[C(\epsilon)]^{1/2}$, whence
\begin{equation}\label{A25}
[C(\epsilon)]^{1/2}=\beta+O(\epsilon)
\end{equation}
and thus
\begin{eqnarray}\label{A26}
R_P(\epsilon)=\int^{\beta+O(\epsilon)}_\beta
\left(\frac{u}{u-\alpha}\right)^{1/2}du\\\nonumber
\approx\int^{\beta+O(\epsilon)}_\beta
\left(\frac{\beta}{u-\alpha}\right)^{1/2}du
\end{eqnarray}
which for all $\beta$ satisfying (\ref{A20}) clearly tends to zero
as $\epsilon\downarrow 0$. Consequently, the ratio of the proper
circumference to the proper radius of $\gamma(\epsilon)$ does not
tend to $2\pi$ as $\epsilon\downarrow 0$, so that there is a
violation of elementary flatness \cite{ref:ER36} at $r=0$ and
$t=t_0$, and since $t_0$ was arbitrary, at $r=0$ for all $t$. This
violation constitutes a ``quasiregular singularity'' \cite{ref:ES77} of
the space-time associated to $g_{PM}$, and thus we can say that
the space-time ($M_0,~g_{PM}$) is analitically inextendible\footnote[7]
{More precisely, there exists no analytic extension in
which $r=0$ corresponds to an interior point in the larger
manifold. Any extension in which $r=0$ corresponds to a
nonpoint-like locus would alter the character of the point mass
and is thus ruled out {\it a priori}.} to $r=0$ for all
$\beta\geq\alpha$. Moreover, since it is well known
\cite{ref:Hagihara31} that those geodesics of the Hilbert metric
$g_H$ that do not run into the boundary at $r^*=\alpha$ are
complete, the diffeomorphism between $g_H$ and $g_{PM}$ for any
admissible $C$ shows that the same is true of those geodesics of
$g_{PM}$ that do not run into the boundary at $r=0$. It follows
that the space-times ($M_0,~g_{PM}$) are maximal.

\section{Acceleration scalar}
As shown by Doughty \cite{ref:Doughty81a}, the acceleration of a
neutral test particle approaching the point mass along a radial
path, as measured by an observer at rest with respect to the point
mass, is given by
\begin{eqnarray}\label{B1}
a=\frac{(-g_{rr})^{1/2}(-g^{rr})|g_{00,r}|}{2g_{00}}
=\frac{A'}{2AB^{1/2}}~for~(5)\\\label{B2}
=\frac{\alpha}{2C(1-\alpha/C^{1/2})^{1/2}}~for~(10)\\\label{B3}
=O[1/2(\alpha r)^{1/2}]~as~r\rightarrow
0~for~(18)
\end{eqnarray}
However, when $R_P(\epsilon)$ in (\ref{A23}) is evaluated for
(\ref{18}), it is readily found that
\begin{equation}\label{B4}
R_P(r)=O[2(\alpha r)^{1/2}]~as~r\rightarrow 0
\end{equation}
thus
\begin{equation}\label{B5}
a\sim1/R_P(r)~as~r\rightarrow 0
\end{equation}
which shows that unlike the Newtonian case, a test particle's
acceleration tends to infinity inversely as the first power of its
proper distance to the central mass.\par

\end{document}